\documentclass[11pt]{article}
\usepackage{amsmath}
\usepackage{amssymb}
\usepackage{amsthm}
\usepackage{amsfonts}
\usepackage{comment}
\usepackage{color}
\usepackage{mathrsfs}
\usepackage{braket}
\usepackage{graphicx}
\usepackage[subrefformat=parens]{subcaption}
\usepackage{cite}
\usepackage{here}

\usepackage[stable]{footmisc}
\makeatletter

\@addtoreset{equation}{section}
\makeatother

\begin{document}

%%%%%%%%%%%%%%%%%%%%%%%%%%
%\maketitle
\pagestyle{empty}

\begin{flushright}
% preprint number
\end{flushright}

\vspace{3cm}

\begin{center}

{\bf\LARGE  
Thermodynamic relation\\on higher-dimensional black hole with arbitrary cosmological constant 
}
\\

\vspace*{1.5cm}
{\large 
Junbeom Ko$^\clubsuit$\footnote{junbum9817@gmail.com}, Bogeun Gwak$^\clubsuit$\footnote{rasenis@dgu.ac.kr}
} \\
\vspace*{0.5cm}

{\it 
$^\clubsuit$Department of Physics, Dongguk University, Seoul 04620, Republic of Korea
}

\end{center}

\vspace*{1.0cm}

\begin{abstract}
{\noindent
We investigated the Goon--Penco (GP) relation on higher-dimensional Reissner-Nordstr\"om black holes with an arbitrary cosmological constant. It was found that the GP relation retained its form in four- and higher-dimensional spacetimes. Thus, the reactions in the black holes are universal with respect to the dimensionality. Furthermore, the GP relation was found to be universal on any state of the black hole including near-extremal and near-Nariai cases. Thus, this study showed that the GP relation was prevalent for higher-dimensional Reissner-Nordstr\"om black holes with an arbitrary cosmological constant regardless of the initial state.
}
\end{abstract}

\newpage
\baselineskip=18pt
\setcounter{page}{2}
\pagestyle{plain}
\baselineskip=18pt
\pagestyle{plain}

\setcounter{footnote}{0}
	
\section{Introduction}
Black holes are treated as massive objects theoretically based on general relativity. In addition, they are predicted to be the end state of the lifetime of stars. Black holes comprise a horizon that has no outgoing geodesics, referred to as the event horizon. No matter can cross the horizon and escape black holes. Inside the horizon, there exists a singularity, which is the endpoint of ingoing geodesics \cite{Penrose:1964wq}. No radiation can be detected owing to these geometric properties. From a classical perspective, black hole mass can be divided into reducible energy and irreducible mass. The reducible energy can be changed through interactions, whereas the irreducible mass is unchangeable \cite{Christodoulou:1970wf,Christodoulou:1971pcn}. Hawking radiation is one of mechanisms through which black holes emit energy in quantum physics. Thus, we can treat black holes as thermal objects that have a certain temperature \cite{Hawking:1974rv,Hawking:1975vcx}. Entropy can be naturally obtained from the surface area of black holes and is referred to as the Hawking--Bekenstein entropy\cite{Bekenstein:1973ur,Bekenstein:1974ax}. Based on establishment of variables, the thermodynamics of black holes were established by Bardeen et al. \cite{Bardeen:1973gs}.

We considered that spacetime has a negative cosmological constant. This is referred to as the anti-de Sitter (AdS) spacetime, which can be defined as a maximally symmetric Lorentzian manifold with negative curvature and a boundary at infinity. This is also depicted as a patch of half of the Einstein static universe. If spacetime has a positive curvature, we define it as the de Sitter(dS) spacetime. Its submanifold is hyperboloid in nature in five-dimensional flat space. Further, its spatial sections contain 3-spheres with positive curvature at fixed $t$. For the AdS spacetime, the theory of gravity on a $D$-dimensional bulk corresponds to the conformal field theory (CFT) on a $D-1$ dimensional boundary. This is referred to as the AdS/CFT correspondence \cite{Maldacena:1997re, Witten:1998qj}. We can expand this concept to dS spacetime such as in AdS spacetime. Then, we can find another correspondence on its boundary named dS/CFT \cite{Strominger:2001gp, Strominger:2001pn}.  

Studies on quantum gravity theory have resulted in the proposal of the weak gravity conjecture (WGC) by Vafa \cite{Vafa:2005ui, Arkani-Hamed:2006emk}. This conjecture explains why gravity is weaker than other gauge forces. This idea can be simply expressed as
    \begin{align}
        \frac{Q}{M}\geq 1
    \label{inq}
    \end{align}
which is saturated on the extremal black hole. In \cite{Banks:2010zn, Harlow:2022ich}, it was reported that this conjecture is based on quantum gravity theory, which does not contain global symmetries. This can be explained through contradictions between global symmetry and in the context of AdS/CFT \cite{Harlow:2022ich}. As evident in the theory of Hawking radiation, black holes do not emit the charged particle itself. Thus, it can be inferred that black holes lose their mass when they emit particles regardless of their global charges. WGC can be considered as a group of agreements to the conjecture. Thus, there are various methods to define the conjecture. Cheung et al. proposed one method of proving the idea by using higher-derivative correction terms on an action to avoid a naked singularity. Thus, the absence of a naked singularity facilitated the verification of the inequality \eqref{inq} \cite{Cheung:2018cwt}. This idea was implemented in other studies \cite{Cheung:2019cwi, Kats:2006xp}, thus facilitating the construction of a relationship between changes in entropy and \eqref{inq}.

WGC can be explained by correction terms \cite{Cheung:2018cwt, Cheung:2019cwi, Kats:2006xp}, which lead to higher-derivative corrections. Goon--Penco mentioned that a derivative in free energy can be evenly applied to an action \cite{Goon:2019faz}. They also defined the corrections with a perturbative parameter that was attached to a cosmological constant. Moreover, corrections to free energy result in changes in thermodynamic variables such as mass, temperature, and entropy. Thus, they provide a relation for the corrected thermodynamic variables after employing the chain rule. This relation can be expanded by applying leading-order expansion to action. Moreover, this approximate relation is related to higher-derivative corrections. Further, the shift in mass is proportional to the change in entropy with a negative sign \cite{Reall:2019sah, Ma:2023qqj}. Thus, we can consider the relation as the proof of WGC. Certain authors have focused on the expansion of application of the relation to other black holes using different variables \cite{Wei:2020bgk, Chen:2020rov, Chen:2020bvv, Sadeghi:2020ciy, Sadeghi:2020xtc}. In \cite{Cano:2019ycn, Ma:2020xwi, McPeak:2021tvu, Noumi:2022ybv, Xiao:2023two}, the authors mentioned a leading-order expansion form for a different case. Other authors have worked on further progressing the relation \cite{Cremonini:2019wdk, McInnes:2021frb, Arkani-Hamed:2021ajd, McInnes:2021zlt, Etheredge:2022rfl}.

In this work, we generalized the Goon--Penco (GP) relation to higher-dimensional Reissner-Nordstr\"om(RN) black holes with an arbitrary cosmological constant. In particular, although the power of gravitational potential changes in the higher-dimensional spacetime, we found that the GP relation is still consistent with that of four dimensions. Thus, the reactions in the black holes are universal with respect to the dimensionality. Furthermore, the universal relation was found to exist in the same form in near-extremal and near-Nariai black holes, and the universal relation was applicable to arbitrary states. This is an extended finding of our previous study 9\cite{Ko:2023nim}. Thus, our results finally showed that the GP relation appeared in higher-dimensional RN black holes with an arbitrary cosmological constant regardless of the initial state.

The remainder of this paper is as organized follows. Section 2 reviews the GP relation. Section 3 examines \eqref{GP} in the context of both five-dimensional Reissner-Nordstr\"om anti-de Sitter(RNAdS) and Reissner-Nordstr\"om anti-de Sitter(RNdS) black holes. Section 4 describes the application of the relation to higher-dimensional RN(A)dS black holes. Finally, Section 5 concludes the paper.

\section{Goon--Penco relation}
\noindent
The GP relation was constructed using mass, temperature, entropy, and perturbation parameter $\epsilon$ \cite{Goon:2019faz}. In addition, each side of the relation includes a partial derivative term of mass and entropy by epsilon,
        \begin{align}
        \frac{\partial M_{ext}(\Vec{\mathcal{Q}},\epsilon)}{\partial\epsilon}=\lim_{M \to M_{ext}(\Vec{\mathcal{Q},\epsilon})} -T\left(\frac{\partial S(M,\Vec{\mathcal{Q}},\epsilon)}{\partial\epsilon}\right)_{M,\Vec{\mathcal{Q}}}
    \label{GP}
    \end{align}
where $M$ and $\Vec{\mathcal{Q}}$ are the mass and additional quantities. This relation can be derived from free energy, $G\left(T, \Vec{\mu}\right)$ using the chain rule. Consequently, we obtain the correction of the cosmological term in action $I$. This correction results in the radius, mass, temperature, and entropy being converted to their corrected forms. Moreover, we performed a leading-order expansion of \eqref{GP}:
    \begin{align}
        \Delta M_{ext}(\Vec{\mathcal{Q}})\approx -T_0(M,\Vec{\mathcal{Q}})\Delta S(M,\Vec{\mathcal{Q}})|_{M\approx M_{ext}^0(\Vec{\mathcal{Q}})}.
    \label{GPapp}
    \end{align}
The expanded form is linked to WGC according to \cite{Cheung:2018cwt}. Naturally, if our calculation satisfies \eqref{GP}, this can facilitate its expansion, i.e., \eqref{GPapp}, and prove WGC is valid on spacetime.

\section{Five-Dimensional RN(A)dS black hole}
\noindent
Before examining the relation in higher dimensions, we consider the solution for the Einstein--Maxwell action in AdS spacetime as a simple example. The action is expressed as
    \begin{align}
        I=-\frac{1}{16\pi}\int d^5x\sqrt{-g}\left(R-F_{\mu\nu}F^{\mu\nu}-2\Lambda\right)
    \end{align}
where $F_{\mu\nu}=\partial_{\mu}A_{\nu}-\partial_{\nu}A_{\mu}$ and $\Lambda=-\frac{6}{l^2}$. However, we added a perturbation parameter to the cosmological constant,
    \begin{align}
        I=-\frac{1}{16\pi}\int d^5x\sqrt{-g}\left(R-F_{\mu\nu}F^{\mu\nu}-2(1+\epsilon)\Lambda\right).
    \label{eq:5a_ac}
    \end{align}
Thus, we obtained a shifted metric as follows:
    \begin{align}
        ds^2=-\frac{\Delta(r)}{r^2}+\frac{r^2}{\Delta(r)}dr^2+r^2d\Omega_3
    \end{align}
where the metric function $\Delta(r)=r^2-2M+\frac{Q^2}{r^2}+\frac{1+\epsilon}{l^2}r^4$ has a perturbative parameter for the cosmological term. We can evaluate thermodynamic variables using this metric as follows:
    \begin{align}
        \begin{split} &
            M_B=\frac{3\Omega_3}{8\pi}M=\frac{3\pi}{4}M, \quad Q_B=\frac{3\Omega_3}{8\pi}Q=\frac{3\pi}{4}Q \\ & T_h=\frac{1}{4\pi}\left(\frac{d\Delta_h}{r_h^2}\right)=\frac{1}{4\pi}\left(\frac{2}{r}+\frac{4(1+\epsilon)r}{l^2}-\frac{2Q^2}{r^5}\right) \\ & S_h=\frac{\Omega_3r_h^3}{4}=\frac{\pi^2r_h^3}{2}
        \end{split}
    \label{eq:5a_tv}
    \end{align}
where $\Omega_3=2\pi^2$ is the surface area of 3-spheres. We define the horizon in terms of entropy as $r_h=\left(\frac{2S_h}{\pi^2}\right)^\frac{1}{3}$. We can calculate the mass parameter from the metric function as $\Delta(r)=0$, expressed as
    \begin{align}
        M=\frac{(2\pi S_h)^{\frac{4}{3}}l^2+\pi^4Q^2l^2+4(1+\epsilon)S_h^2}{2^{\frac{5}{3}}\pi^{\frac{8}{3}}S_h^{\frac{2}{3}}l^2}.
    \label{eq:5a_M}
    \end{align}
Using \eqref{eq:5a_M}, we build the epsilon function as
    \begin{align}
        \epsilon=-\frac{l^2\left(\pi^4Q^2-2^\frac{5}{3}\pi^\frac{8}{3}S_h^\frac{2}{3}M+(2S_h)^\frac{4}{3}\right)}{4S_h^2}-1.
    \label{eq:5a_ep}
    \end{align}
Consequently, we consider a partial derivative of \eqref{eq:5a_ep} to obtain the left-hand side of the GP relation,
    \begin{align}
        \frac{\partial\epsilon}{\partial S_h}=\frac{l^2(\pi^4Q^2-(2\pi S_h)^\frac{4}{3})-8S_h^2(1+\epsilon)}{6S_h^3}.
    \label{eq:5a_pa}
    \end{align}
Subsequently, we obtain the temperature using \eqref{eq:5a_tv},
    \begin{align}
        T=-\frac{l^2\left(\pi^4Q^2-(2\pi S_h)^\frac{4}{3}\right)-8S_h^2(1+\epsilon)}{2^\frac{8}{3}\pi^\frac{5}{3}S_h^\frac{5}{3}l^2}.
    \label{eq:5a_T}
    \end{align}
We can build the right-hand side of the relation by combining the inverse of \eqref{eq:5a_pa} as
    \begin{align}
        -T\frac{\partial S_h}{\partial\epsilon}=\frac{3S_h^\frac{4}{3}}{2^\frac{5}{3}\pi^\frac{5}{3}l^2}.
    \label{eq:5a_rh}
    \end{align}
For the left-hand side of the relation, we consider the derivative of $M_B$ in terms of $\epsilon$,
    \begin{align}
        \frac{\partial M_B}{\partial \epsilon}=\frac{3S_E^\frac{4}{3}}{2^\frac{5}{3}\pi^\frac{5}{3}l^2}.
    \label{eq:5a_lf}
    \end{align}
Further, we substitute $T=0$ to determine the extremality condition,
    \begin{align}
        \begin{split} &
            S_E=\left(-\frac{\pi^\frac{4}{3}\left(l^4+l^2P^\frac{1}{3}+P^\frac{2}{3}\right)}{6\times2^\frac{2}{3}(1+\epsilon)P^\frac{1}{3}}\right)^\frac{3}{2} \\ &
            P=l^6-54Q^2l^2(1+\epsilon)^2+6\sqrt{3}\sqrt{-Q^2l^4(1+\epsilon)^2(l^4-27Q^2(1+\epsilon)^2)}.
        \end{split}
    \label{eq:5a_ext}
    \end{align}
Evidently, \eqref{eq:5a_rh} and \eqref{eq:5a_lf} have the same form. Moreover, we can construct an equation relating them and substitute \eqref{eq:5a_ext} as follows:
    \begin{align}
        \frac{\partial M_B}{\partial\epsilon}= \lim_{M \to M_{ext}}\left(-T\frac{\partial S}{\partial\epsilon}\right)_{M,Q}=\frac{3S_E^\frac{4}{3}}{2^\frac{5}{3}\pi^\frac{5}{3}l^2}.
    \label{eq:5a_GP}
    \end{align}
As evident in \eqref{eq:5a_GP}, the relation appears valid for five-dimensional RNAdS black holes when we change the mass to real mass. Notably, the relation can be expanded to five-dimensional spacetime. Further, it can be applied to the Nariai limit in the same case.

Next, we calculate this relation for five-dimensional charged black holes with a positive cosmological constant, that is, RNdS black holes. The action is conveyed as changing the sign of the cosmological term in \eqref{eq:5a_ac} into $+2(1+\epsilon)\Lambda$. Thus, the metric function can be expressed as
    \begin{align}
        \Delta(r)=r^2-2M+\frac{Q^2}{r^2}-\frac{1+\epsilon}{l^2}r^4.
    \end{align}
Consequently, the variables become
    \begin{align}
        \begin{split} &
            M=\frac{r^2}{2}+\frac{Q^2}{2r^2}-\frac{1+\epsilon}{2l^2}r^4=\frac{(2\pi S_h)^\frac{4}{3}+\pi^4Q^2l^2-4(1+\epsilon)S_h^2}{2^\frac{5}{3}\pi^\frac{8}{3}S_h^\frac{2}{3}l^2} \\ &
            T_h=\frac{1}{4\pi}\left(\frac{2}{r}-\frac{4(1+\epsilon)r}{l^2}-\frac{2Q^2}{r^5}\right)=-\frac{l^2\left(\pi^4Q^2-(2\pi S_h)^\frac{4}{3}\right)+8S_h^2(1+\epsilon)}{2^\frac{8}{3}\pi^\frac{5}{3}S_h^\frac{5}{3}l^2},
        \end{split}
    \label{eq:5d_va}
    \end{align}
with horizon, $r_h=\left(\frac{2S_h}{\pi^2}\right)^\frac{1}{3}$. In contrast, the entropy has the same form as that of the AdS case, that is, $S_h=\frac{\pi^2r_h^3}{2}$. Similar to the RNAdS case, we construct the epsilon function using the mass parameter $M$,
    \begin{align}
        \epsilon=\frac{l^2\left(\pi^2Q^2-2^\frac{5}{3}\pi^\frac{8}{3}S_h^\frac{2}{3}M+(2\pi S_h)^\frac{4}{3}\right)}{4S_h^2}-1.
    \label{eq:5d_ep}
    \end{align}
Subsequently, we consider the partial derivative of \eqref{eq:5d_ep} with $S_h$,
    \begin{align}
        \frac{\partial \epsilon}{\partial S_h}=\frac{-l^2\left(\pi^4Q^2-(2\pi S_h)^\frac{4}{3}\right)-8S_h^2(1+\epsilon)}{6S_h^3}.
    \label{eq:5d_paep}
    \end{align}
Next, we combine the temperature in \eqref{eq:5d_va} and the inverse of \eqref{eq:5d_paep} to obtain
    \begin{align}
        -T\frac{\partial S_h}{\partial \epsilon}=-\frac{3S^\frac{4}{3}}{2^\frac{5}{3}\pi^\frac{5}{3}l^2}
    \label{eq:5d_rh}
    \end{align}
which can be the base of the right-hand side of the GP relation. We construct the left-hand side of the GP relation with real mass and the perturbation parameter,
    \begin{align}
        \frac{\partial M_B}{\partial \epsilon}=-\frac{3S^\frac{4}{3}}{2^\frac{5}{3}\pi^\frac{5}{3}l^2}.
    \label{eq:5d_lf}
    \end{align}
Consequently, we obtain the extremality condition for two types of limits: the extremal and Nariai limits\cite{Zhen:2024bcp},
    \begin{align}
        \begin{split} &
            S_{E,N}=\left(\frac{\pi^\frac{4}{3}\left(l^4+l^2P^\frac{1}{3}+P^\frac{2}{3}\right)}{6\times2^\frac{2}{3}(1+\epsilon)P^\frac{1}{3}}\right)^\frac{3}{2} \\ &
            P=l^6-54l^2Q^2(1+\epsilon)^2+6\sqrt{3}\sqrt{-l^4Q^2(1+\epsilon)^2(l^4-27Q^2(1+\epsilon)^2)}.
        \end{split}
    \label{eq:5d_ext}
    \end{align}
We can choose the Nariai condition in five-dimensional RNdS black holes through comparisons with the same condition in the Schwarzschild case, $S_N=\left(\frac{\pi^\frac{4}{3}l^2}{2^\frac{5}{3}(1+\epsilon)}\right)^\frac{3}{2}$. The positive case in \eqref{eq:5d_ext} can be the Nariai condition, whereas the other is the extremal condition of five-dimensional RNdS black holes. If we substitute the condition in \eqref{eq:5d_ext} into \eqref{eq:5d_rh} and \eqref{eq:5d_lf}, we find that the relation $\frac{\partial M_B}{\partial\epsilon}=\lim_{M \to M_{ext}}-T\frac{\partial S_{E,N}}{\partial\epsilon}$ is also valid.

\section{Higher dimensional RN(A)dS black holes : General case}
\noindent
Next, we apply our calculation to a more general case, that is, higher-dimensional black holes. According to our results in the previous section, it should work well. The action of higher dimensional AdS black holes is expressed as
    \begin{align}
        I=-\frac{1}{16\pi}\int d^Dx\sqrt{-g}\left(R-F_{\mu\nu}F^{\mu\nu}-2(1+\epsilon)\Lambda\right)
    \end{align}
where $D$ is the dimension of spacetime. Thus, the metric function is expressed as
    \begin{align}
        \Delta(r)=r^2-\frac{2M}{r^{D-5}}+\frac{Q^2}{r^{2D-8}}+\frac{1+\epsilon}{l^2}r^4.
    \end{align}
We can obtain thermodynamic variables as follows:
    \begin{align}
        \begin{split} &
             M=\frac{1}{2}\left(r_h\right)^{D-3}+\frac{1+\epsilon}{2l^2}\left(r_h\right)^{D-1}+\frac{Q^2}{2}\left(r_h 
             \right)^{3-D} \\ &
             T_h=\frac{1}{4\pi}\left(\frac{D-3}{r_h}-\frac{(D-3)Q^2}{\left(r_h\right)^{2D-5}}+\frac{(D-1)r_h}{l^2}\right) \\ &
             S_h=\frac{\Omega_{D-2}\left(r_h\right)^{D-2}}{4}.
        \end{split}
    \label{eq:ha_TV}
    \end{align}
We derive the function of epsilon from mass using \eqref{eq:ha_TV} and substitute $r_h=\left(\frac{4S_h}{\Omega_{D-2}}\right)^\frac{1}{D-2}$, where $\Omega_{D-2}$ is surface area of $(D-2)$-sphere.
    \begin{align}
        \epsilon=2Ml^2\left(\frac{4S_h}{\Omega_{D-2}}\right)^\frac{1-D}{D-2}-Q^2l^2\left(\frac{4S_h}{\Omega_{D-2}}\right)^{-2}-l^2\left(\frac{4S_h}{\Omega_{D-2}}\right)^\frac{-2}{D-2}-1.
    \label{eq:ha_ep}
    \end{align}
Consequently, we consider the partial derivative of \eqref{eq:ha_ep} with $S_h$ as follows:
    \begin{multline}
        \frac{\partial \epsilon}{\partial S
        }=-\frac{4l^2}{(D-2)\Omega_{D-2}}\left(\frac{4S_h}{\Omega_{D-2}}\right)^\frac{1-D}{D-2}\times \\ \left((D-3)\left(\frac{4S_h}{\Omega_{D-2}}\right)^\frac{-1}{D-2}-Q^2(D-3)\left(\frac{4S_h}{\Omega_{D-2}}\right)^\frac{-2D+5}{D-2} +\frac{(1+\epsilon)(D-1)}{l^2}\left(\frac{4S_h}{\Omega_{D-2}}\right)^\frac{1}{D-2}\right).
    \label{eq:ha_paep}
    \end{multline}
Thus, the temperature is obtained as
    \begin{multline}
        T_h=\frac{1}{4\pi}\times \\ \left((D-3)\left(\frac{4S_h}{\Omega_{D-2}}\right)^\frac{-1}{D-2}-Q^2(D-3)\left(\frac{4S_h}{\Omega_{D-2}}\right)^\frac{-2D+5}{D-2}+\frac{(1+\epsilon)(D-1)}{l^2}\left(\frac{4S_h}{\Omega_{D-2}}\right)^\frac{1}{D-2}\right).
    \label{eq:ha_T}
    \end{multline}
Further, we can build the base of the right-hand side of the relation as
    \begin{align}
        -T\frac{\partial S}{\partial \epsilon}=\frac{(D-2)\Omega_{D-2}}{16\pi l^2}\left(\frac{4S_h}{\Omega_{D-2}}\right)^\frac{D-1}{D-2}.
    \label{eq:ha_rh}
    \end{align}
We construct the left-hand side by considering the partial derivative of the real mass such as that for five-dimensional RN(A)dS black holes as follows:
    \begin{align}
        \frac{\partial M_B}{\partial \epsilon}=\frac{(D-2)\Omega_{D-2}}{16\pi l^2}\left(\frac{4S_h}{\Omega_{D-2}}\right)^\frac{D-1}{D-2}.
    \label{eq:ha_lf}
    \end{align}
The extremality condition varies depending on the dimension $D$. Therefore, we can only derive the general form using \eqref{eq:ha_rh} and \eqref{eq:ha_lf}. This facilitates the construction of the fundamental form of the relation as $\frac{\partial M_B}{\partial \epsilon}=-T\frac{\partial S}{\partial \epsilon}=\frac{(D-2)\Omega_{D-2}}{16\pi l^2}\left(\frac{4S_h}{\Omega_{D-2}}\right)^\frac{D-1}{D-2}$. We can also reproduce the result of five-dimensional RN(A)dS black holes if we set $D=5$.

Next, we consider the case of a positive cosmological constant, dS spacetime. In contrast to that in the AdS case, the metric function is
    \begin{align}
        \Delta(r)=r^2-\frac{2M}{r^{D-5}}+\frac{Q^2}{r^{2D-8}}-\frac{1+\epsilon}{l^2}r^4.
    \label{eq:hd_mf}
    \end{align}
Then, we can build variables using \eqref{eq:hd_mf},
    \begin{align}
        \begin{split} &
            M=\frac{1}{2}\left(r_h\right)^{D-3}+\frac{1+\epsilon}{2l^2}\left(r_h\right)^{D-1}+\frac{Q^2}{2}\left(r_h\right)^{3-D} \\ &
            T_h=\frac{1}{4\pi}\left(\frac{D-3}{r_h}-\frac{(D-3)Q^2}{\left(r_h\right)^{2D-5}}+\frac{(D-1)r_h}{l^2}\right) \\ &
            S_h=\frac{\Omega_{D-2}\left(r_h\right)^{D-2}}{4}.
        \end{split}
    \label{eq:hd_TV}
    \end{align}
From \eqref{eq:hd_TV}, we can obtain the epsilon from mass, $M$, with $r_h=\left(\frac{4S_h}{\Omega_{D-2}}\right)^\frac{1}{D-2}$,
    \begin{align}
        \epsilon=-2Ml^2\left(\frac{4S_h}{\Omega_{D-2}}\right)^{\frac{1-D}{D-2}}+l^2\left(\frac{4S_h}{\Omega_{D-2}}\right)^{\frac{-2}{D-2}}+Q^2l^2\left(\frac{4S_h}{\Omega_{D-2}}\right)^{-2}.
    \label{eq:hd_ep}
    \end{align}
Now, we can build \eqref{GP} using the partial derivative term of $M$ and $S$. Starting with the right side of the relation, we calculate derivative of entropy as follows:
    \begin{align}
        \begin{split} &
            \frac{\partial \epsilon}{\partial S}=\frac{4l^2}{(D-2)\Omega_{D-2}}\left(\frac{4S_h}{\Omega_{D-2}}\right)^{\frac{1-D}{D-2}}\times \\ &
            \left((D-3)\left(\frac{4S_h}{\Omega_{D-2}}\right)^\frac{-1}{D-2}-(D-3)Q^2\left(\frac{4S_h}{\Omega_{D-2}}\right)^\frac{5-2D}{D-2}-\frac{(1+\epsilon)(D-1)}{l^2}\left(\frac{4S_h}{\Omega_{D-2}}\right)^\frac{1}{D-2}\right).
        \end{split}
    \label{eq:hd_paep}
    \end{align}
Next, we obtain the temperature as follows:
    \begin{multline}
        T_h=\frac{1}{4\pi}\times \\ 
        \left((D-3)\left(\frac{4S_h}{\Omega_{D-2}}\right)^\frac{-1}{D-2}-(D-3)Q^2\left(\frac{4S_h}{\Omega_{D-2}}\right)^\frac{5-2D}{D-2}-\frac{(1+\epsilon)(D-1)}{l^2}\left(\frac{4S_h}{\Omega_{D-2}}\right)^\frac{1}{D-2}\right).
    \label{eq:hd_T}
    \end{multline}
By combining \eqref{eq:hd_paep} and \eqref{eq:hd_T}, we can build the form of the right-hand side,
    \begin{align}
        -T\frac{\partial S}{\partial \epsilon}=-\frac{(D-2)\Omega_{D-2}}{16\pi l^2}\left(\frac{4S_h}{\Omega_{D-2}}\right)^\frac{D-1}{D-2}.
    \label{eq:hd_rh}
    \end{align}
In the case of higher-dimensional black holes, the left side of the relation becomes a partial derivative of the real mass $M_B=\frac{(D-2)\Omega_{D-2}}{8\pi}$,
    \begin{align}
        \frac{\partial M_B}{\partial \epsilon}=-\frac{(D-2)\Omega_{D-2}}{16\pi l^2}\left(\frac{4S_h}{\Omega_{D-2}}\right)^\frac{D-1}{D-2}.
    \label{eq:hd_lf}
    \end{align}
Thus, we can consider a general case of \eqref{GP} by checking whether the results of \eqref{eq:hd_rh} and \eqref{eq:hd_lf} are the same, that is, $\frac{\partial M_B}{\partial \epsilon}=-T\frac{\partial S}{\partial \epsilon}=-\frac{(D-2)\Omega_{D-2}}{16\pi l^2}\left(\frac{4S_h}{\Omega_{D-2}}\right)^\frac{D-1}{D-2}$. Thus, the GP relation can be generally applied to higher dimensions. This can be considered as the generalized form of the GP relation.

\section{Summary}
This study investigated the universality of the GP relation on RN black holes with positive and negative cosmological constants. By considering a higher dimension $D$, we constructed a generalized form of the relation. This was in line with the results of a previous study \cite{Ko:2023nim} when considering the $D=4$ limit. We can verify that the WGC is applicable to arbitrary dimensional RN(A)dS black holes by building a GP relation on five- or higher-dimensional black holes such as in \cite{Goon:2019faz}.

For examining the application of our calculation to five- or higher-dimensional RN(A)dS black holes, first, we constructed thermodynamic variables using a metric function, $\Delta(r)$, for five-dimensional RNAdS black holes. We formulated a relation as earlier. The relation was constructed appropriately. We reaffirmed that the relation was valid on five-dimensional RNAdS black holes. We also applied our formulation to RNdS black holes. Upon substituting \eqref{eq:5d_ext} into \eqref{eq:5d_rh} and \eqref{eq:5d_lf}, we found that the relation fit well in case of the 5-dimensional RNdS black holes. Subsequently, we introduced higher-dimensional RN(A)dS black holes. However, we could not calculate the extremality condition because the dimension could not be determined. As presented in Sec. 4, \eqref{eq:ha_rh} and \eqref{eq:ha_lf} were same for the RNAdS cases. Thus, the relation can be valid in its generalized form. By confirming that \eqref{eq:hd_rh} and \eqref{eq:hd_lf} are the same, we can verify the generalization for RNdS black holes as well.

We expanded our calculations to higher-dimensional black holes. The relation was revealed as being more universal in nature. The proportional relation between the shifted mass, temperature, and entropy provides insights into the WGC. Thus, the results of this study can facilitate a deeper understanding of quantum gravity. Further, these results are expected to provide further motivation for future studies in this field.

\vspace{10pt} 

\noindent{\bf Acknowledgments}

\noindent This research was supported by Basic Science Research Program through the National Research Foundation of Korea (NRF) funded by the Ministry of Education (NRF-2022R1I1A2063176) and the Dongguk University Research Fund of 2024. BG appreciates APCTP for its hospitality during completion of this work.\\

\bibliographystyle{jhep}
\bibliography{ref}
\end{document}